\documentclass[%
 aps,
 prb,
 amsmath,amssymb,
 reprint,
]{revtex4-2}

\usepackage{graphicx}
\usepackage{dcolumn}
\usepackage{bm}

\usepackage[utf8]{inputenc}
\usepackage[T1]{fontenc}
\usepackage{etoolbox}
\usepackage{float}

\begin{document}

\title{Predicting New Heavy Fermion Materials within Carbon-Boron Clathrate Structures}
\author{Rishi Rao}
\author{Li Zhu}%
\email{li.zhu@rutgers.edu}
\affiliation{ 
Department of Physics, Rutgers University, Newark, New Jersey 07102, USA
}%

\date{\today}

\begin{abstract}  
     Heavy fermion materials have been a rich playground for strongly correlated physics for decades. However, engineering tunable and synthesizable heavy fermion materials remains a challenge. We strive to integrate heavy fermion properties into carbon boron clathrates as a universal structure which can host a diverse array of interesting physical phenomena. Using a combination of density functional theory and dynamical mean field theory, we study two rare earth carbon boron clathrates, SmB$_3$C$_3$ and CeB$_3$C$_3$, and explore properties arising from the strong electronic correlations. We find a significant increase in the density of states at the Fermi level in CeB$_3$C$_3$ as the temperature is lowered, indicating the development of a heavy electron state. In SmB$_3$C$_3$, a potential Kondo insulating state is identified. Both findings point to rare earth carbon boron clathrates as novel strongly correlated materials within a universally tunable structure, offering a fresh platform to  innovate upon conventional heavy-fermion materials design.
\end{abstract}

\maketitle

\section{Introduction}

Heavy-fermion materials represent a fascinating class of quantum systems with nearly half century of dedicated research \cite{ghfWirth2016}. Characterized by extremely large effective electron masses due to strong electron-electron interactions, these materials offer a rich playground for exploring exotic quantum states of matter, including unconventional superconductivity \cite{ghfWhite2015,ghfNorman2011,ghfSmidman2018,ghfMathur1998}, quantum criticality\cite{ghfGegenwart2008, ghfStockert2011,ghfAynajian2012}, and non-Fermi liquid behavior \cite{ghfColeman1999,ghfSeiro2018,ghfLohneysen1994,ghfLohneysen1995}. 
While there have been significant advancements in the field of heavy fermions over the decades, persistent challenges and open questions still invite further exploration. Central to this pursuit is deepening our understanding of existing materials and exploring the potential of novel systems.  Through the development of new heavy-fermion systems, we can delve deeper into the open questions, advancing both our scientific understanding and opening avenues for technological applications. 

The growth in computational capacities and algorithmic sophistication in recent years has accelerated materials science research. However, progress in predicting and discovering new heavy-fermion materials remains incremental \cite{ghfThompson2012, ghfNaritsuka2021, PhysRevB.30.1249, svanidzeEmpiricalWayFinding2019, tangSearchingNewHeavy1989,fiskSearchingHeavyFermion2006,ghfShishido2010,ghfJang2022}. Most findings are anchored within a subset of metallic compounds. Conventionally, heavy-fermion materials are exemplified by a limited category of compounds that contain rare-earth ions with partially filled $f$-shells \cite{ghfColeman2015}. These materials exhibit heavy-fermion properties but are limited by a lack of structural diversity, with little structural connection between compounds containing different rare earth ions. Recently, a surge of interest has been directed towards two-dimensional (2D) heavy-fermion materials, offering a new approach to this field. For example, 2D CeIn$_3$ was  produced through artificial superlattices \cite{ghfShishido2010}, in which electronic properties display striking deviations from the standard low-temperature Fermi liquid behavior. CeSiI was also predicted to be a 2D van der Waals heavy-fermion system through a data mining theoretical approach \cite{ghfJang2022}. Nevertheless, many of these are derived from the bulk form of pre-existing heavy-fermion compounds. While these materials present a new avenue for exploration, they fail to address the need for greater structural diversity and selective tunability. In addition, it is of interest to create standardized structures where tuning can be applied in a systematic manner while limiting other degrees of freedom. Thus, there is still a pressing need for new material platforms that can be tailored to exhibit either heavy-fermion behavior or other properties of interest in a consistent and adjustable manner.

In this work, we propose a strategy to address these challenges by leveraging the properties of sodalite Carbon-Boron (C-B) clathrates. These unique frameworks, which have been successfully synthesized in the SrB$_3$C$_3$ \cite{cbZhu2020} and LaB$_3$C$_3$ \cite{cbStrobel2021} systems, offer a fresh perspective on the heavy-fermion landscape. The versatility of the clathrate structure allows for targeted doping strategies, and the unique cage-like architecture provides a controlled environment for manipulating electron-electron interactions. The electronic structure of C-B clathrates can be altered immensely by encapsulating guest dopant atoms within the host cavities. For example, Sc-doped C-B clathrate exhibit excellent ferroelectric \cite{zhuPredictionExtendedFerroelectric2020} and piezoelectric properties \cite{cbZhi2023}, and doped C-B clathrates have been identified as superconductors with high critical temperatures \cite{cbNisha2023, cbZhu2023, PhysRevB.105.224514, PhysRevB.105.094503, PhysRevB.105.064516, PhysRevB.103.144515}.
By selectively introducing rare-earth ions ($e.g.$, lanthanoids) within the cages, the confined space of C-B framework could effectively enhance the interplay between localized $f$-electrons and itinerant conduction electrons, a key ingredient for heavy-fermion behavior, making C-B clathrates an attractive system in the design of heavy-fermion materials.
Furthermore, the successful synthesis of LaB$_3$C$_3$ provides a compelling precedent for the viability of other lanthanoids within the C-B clathrate framework.

In this study, guided by state-of-the-art first-principles calculations, we have identified two promising candidates as model systems that have the potential to exhibit heavy-fermion behavior within the clathrate architecture: CeB$_3$C$_3$ and SmB$_3$C$_3$. These two model systems provide a fresh perspective into a strongly correlated landscape within the C-B clathrate platform, laying the foundation for future studies and potentially opening up a new chapter in the study of quantum materials within the clathrate framework.

\section{Methods}
The electronic properties were computed through a combination of density functional theory with dynamical mean field theory (DFT+DMFT) as implemented within the COMDMFT software package \cite{comdmftKutepov2017}. The Kohn-Sham DFT equations were solved using a Linear Augmented Plane Wave method as implemented within the Flapwmbpt code \cite{comdmftKutepov2017}. The effective hamiltonian was constructed from Wannier orbitals using the Wannier90 package \cite{comdmftArash2008}. Coulomb repulsion and Hund's coupling terms were  obtained through the constrained random phase approximation (cRPA) method \cite{Amadon2014} using the ABINIT software framework \cite{Gonze2020,Romero2020}. In particular, the values $U=$ 4.67 eV and $J=$ 0.3 eV were used for Ce$_2$B$_2$C$_2$ while $U=$ 4.95 eV and $J=$ 0.4 eV were used for Sm$_2$B$_2$C$_2$. The same values of $U$ and $J$ were used for all temperatures in the DMFT calculations. Slater parametrization is used for generating the Coulomb interaction terms. Continuous time quantum Monte Carlo \cite{ctqmcGull2011} was used as a solver for the impurity problem and calculations were iterated until charge self-consistency was achieved with double counting being handled through the nominal double counting scheme \cite{comdmftHaule2010}. All calculations were carried out in the paramagnetic phase with spin-orbit coupling included. Phonon calculations with spin-orbit coupling were carried out using the Phonopy package \cite{phonopy1,phonopy2} combined with the Vienna $ab$ $initio$ simulation package (VASP) code \cite{VASP}. 
See Supplemental Material \cite{supplementlink} (see also \cite{Torrent2008,Gonze2020,Romero2020,Amadon2014,ceriumNikolaev2012,samariumEllinger1953,boronOganov2009,carbonLindsay1927} therein) for more details on computational methods.

\begin{figure}
\includegraphics[width=0.3\textwidth]{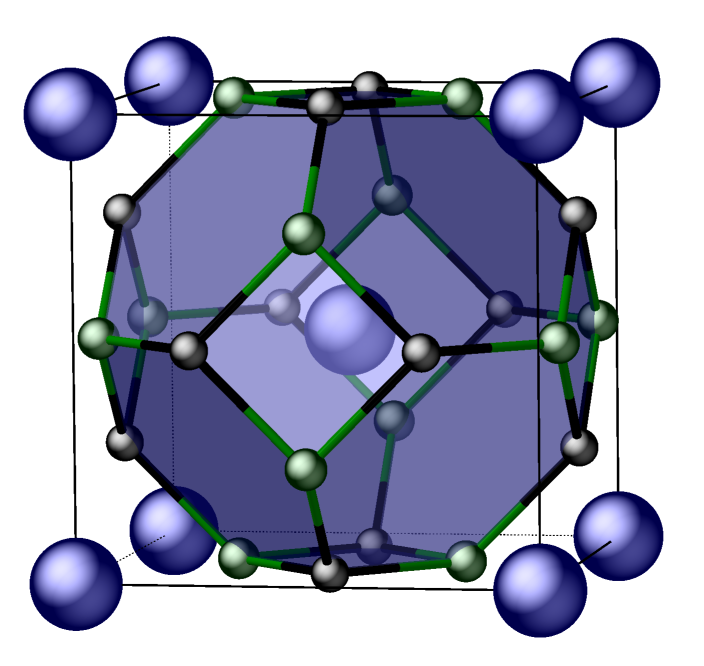}
\caption{The crystal structure of the $Pm\bar{3}n$ carbon-boron clathrate. The blue, black, and green spheres represent the Ce/Sm, C, and B atoms, respectively.} 
\label{fig:Ce2Structure}
\end{figure}


Our first model system centers on the CeB$_3$C$_3$ clathrate, which was constructed based on the structure of the recently synthesized SrB$_3$C$_3$/LaB$_3$C$_3$. Ce occupies a prominent position in the field of heavy fermion research due to its small 4$f$ magnetic moment, which can be easily screened by conduction electrons \cite{ce2Galler2021}. This prominence traces back to the discovery of heavy-fermion behavior in CeAl$_3$ \cite{ce2Andres1975}, and it is responsible for the strongly correlated physics in a wide variety of heavy fermion systems \cite{ce2Flouquet1990, ce2Matsumoto2010,ce2Zhou2018}. 
Therefore, incorporating Ce into sodalite C-B clathrate structures is likely to establish a promising foundation for exploring heavy fermion behavior. The absence of imaginary frequencies in the phonon dispersion spectrum of CeB$_3$C$_3$ (See supplemental material Fig. S1(a) \cite{supplementlink}) suggests structural stability and a negative predicted formation energy (approximately -2.03 eV per formula unit) indicates synthesizability under ambient conditions.

\section{RESULTS AND DISCUSSION}
\begin{figure}
    \includegraphics[width=0.5\textwidth]{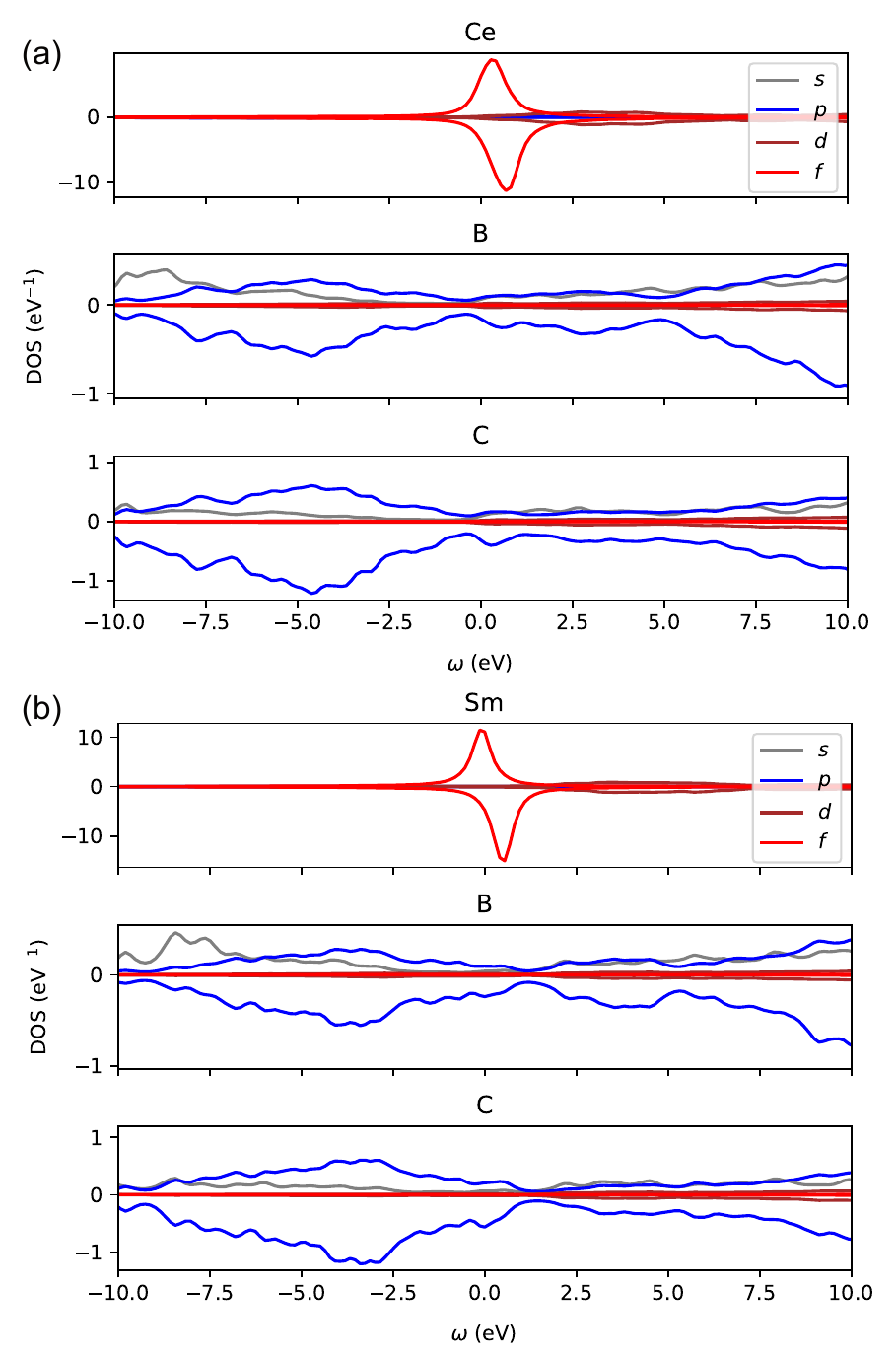}
    \caption{Projected density of states for (a) CeB$_3$C$_3$ and (b) SmB$_3$C$_3$ calculated at the DFT level. The Fermi level is set at 0 eV.}
    \label{fig:dft_pdos}
\end{figure}

We initiate our investigation by examining the ground state electronic structure of CeB$_3$C$_3$, as the Kondo effect stems from the screening of local magnetic moments due to hybridization between conduction and impurity electrons. We find the 2$p$ electrons of B and C play a dominant role in this hybridization due to a strong overlap in the density of states near the fermi level at the LDA level as shown in Fig. \ref{fig:dft_pdos}(a). 
Anisotropy can also significantly affect the properties and utility of heavy-fermion materials \cite{ce2Moll2017}. In order to classify the anisotropy, we examine the contributions from various orbitals to the ground state. Spin-orbit coupling splits the degenerate 4$f$ orbital states into $j=5/2$ and $7/2$ angular momentum components, with the 4$f_{5/2}$ being the ground state. The cubic crystal field further removes the degeneracy of the six 4$f_{5/2}$ states and splits them into a $\Gamma_8$ quartet and a $\Gamma_7$ doublet. These states are separated by 14 K, with the lowest energy state being $\Gamma_8$, as is typical for Ce compounds in cubic crystal fields \cite{ce2Souma2001}. Since splitting is not significant, its energy scale may match that of the Ruderman–Kittel–Kasuya–Yosida (RKKY) interaction at very low temperatures and enhance formation of long-range magnetic order \cite{ce2Konic2023}.

\begin{figure*}
    \centering
    \includegraphics[width=0.8\textwidth]{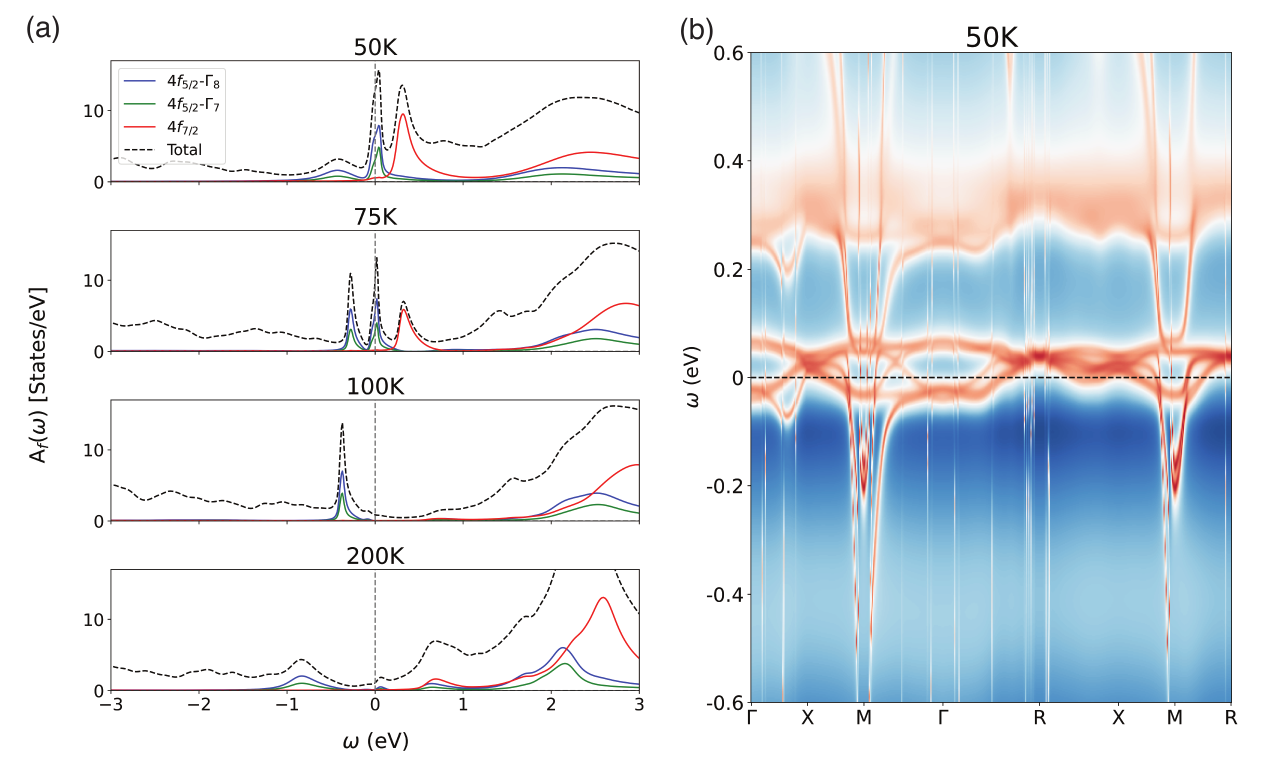}
    \caption{Temperature evolution of Ce 4$f$-electron density of states and spectral function at 50K. (a) Temperature evolution of the 4$f$-electron density of states of CeB$_3$C$_3$. (b)  Spectral function A($\vec{k}, \omega$) of CeB$_3$C$_3$ at T = 50 K. }
    \label{fig:Ce2tempevo}
\end{figure*}

Since we are investigating a Kondo lattice system, it is crucial to examine the behavior of the 4$f$ electron states near the Fermi level to deduce the onset of coherent scattering and classify the observed properties. In particular, an analysis tracing the evolution of the $f$-electron density of states from high to low temperatures can show the development of coherence. As shown in Fig. \ref{fig:Ce2tempevo}(a), the Ce $f$ electrons transition from a local moment phase, exhibiting no spectral weight at the Fermi level, to a delocalized metallic state with a large spectral weight at the Fermi level as the temperature decreases from 200 K to 50 K. 
At 200 K, the lower and upper Hubbard bands are located around -0.9 eV and 2.5 eV, respectively, with shifts observed when transitioning between local moment and itinerant phases. The concept of Hubbard bands is applicable in this case since dynamical mean field theory maps the Hubbard Model onto a single impurity Anderson model in the limit of infinite dimensions \cite{ghfGeorges1992}. Therefore, the splitting observed in the Hubbard model, due to on-site electron interactions, is also present in the impurity model treated here.

The spectral weight gradually redistributes away from the lower and upper Hubbard bands as the peak at the Fermi level becomes more resolved. The splitting due to spin-orbit coupling is readily apparent, with the $j=7/2$ states being elevated above the $j=5/2$ states by around 276 meV at 50 K. The evolution observed aligns well with the two fluid model \cite{ce2Yang2016}. At high temperatures, the lower and upper Hubbard bands, comprised of the 4$f$ electron states, contain almost all of the spectral weight. This represents a localization of the Ce electrons and results in the suppression of the Kondo peak. However, as the temperature is lowered, Ce electrons begin to hybridize with the $p$ electrons of B and C atoms, leading to the formation of itinerant quasiparticles near the Fermi level and a Kondo resonance peak. In Fig. \ref{fig:Ce2tempevo}(a), we can observe that the spectral weight is rapidly shifted toward the Fermi level when the temperature drops below 100 K, moving closer to the peak of the Fermi level. The low temperature results are qualitatively similar to the model calculations from Bickers \textit{et al.} \cite{PhysRevB.36.2036}, where a simple effective Hamiltonian is treated using a large-degeneracy expansion and taking into account spin-orbit coupling. At 50 K, a three-peak structure is present with the ground multiplet at -0.4 eV, a Kondo resonance just above the Fermi level, and an excited peak around 0.276 eV. The clear presence of the Kondo resonance peak indicates substantial hybridization between the Ce $f$ electrons and the conduction electrons, suggesting the formation of a heavy fermion state. This resonance indicates a significant admixture of $f^0$ and $f^1$ states in the interacting ground state, as expected for coherent scattering off of Ce 4$f$ moments. 

\begin{figure}
    \includegraphics[width=0.5\textwidth]{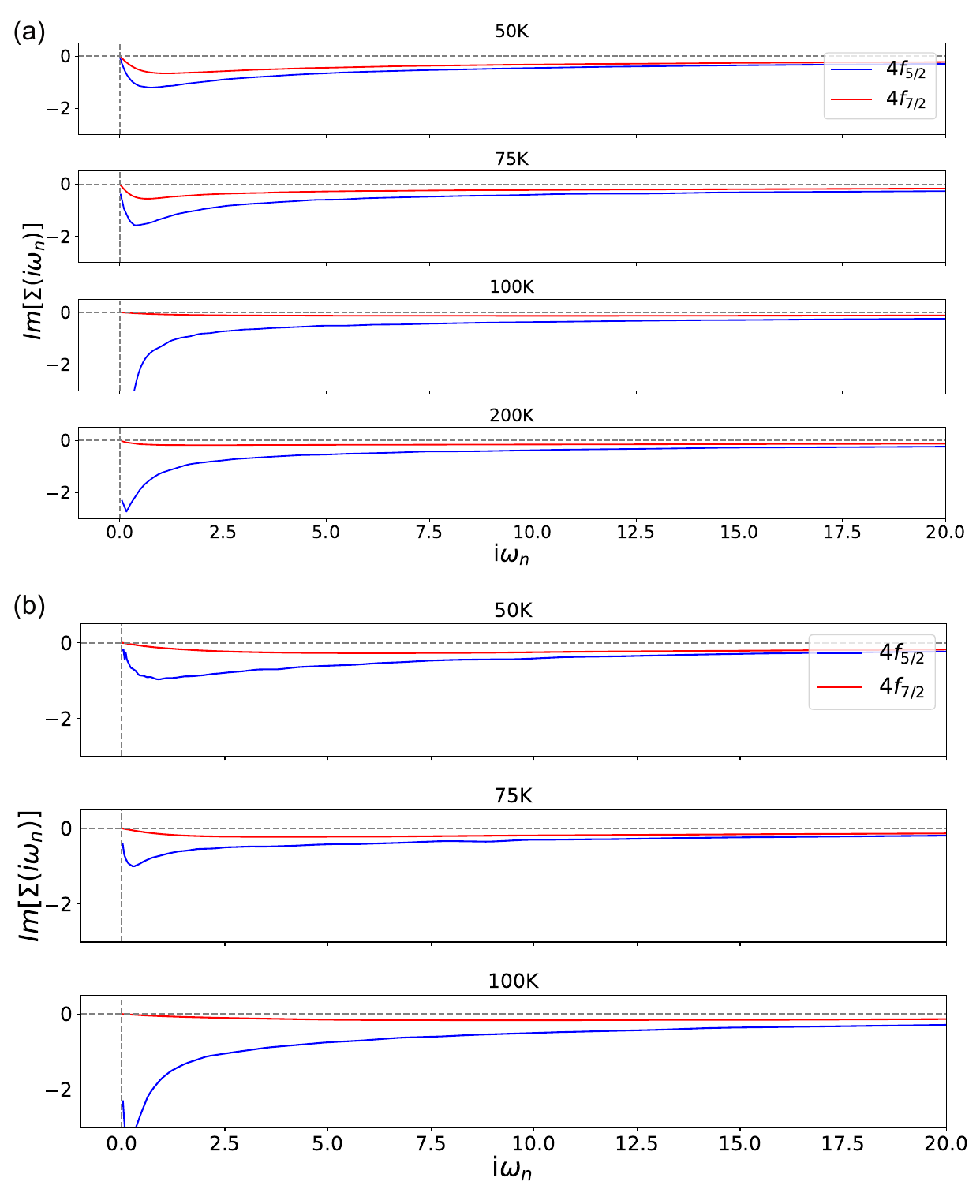}
    \caption{Temperature evolution of imaginary part of the 4$f$ electron self energy for each angular momentum state on the Matsubara axis for (a) CeB$_3$C$_3$ and (b) SmB$_3$C$_3$. }
    \label{fig:imsigtempevo}
\end{figure}

The observed behavior near the Fermi level lends itself to analysis via Landau-Fermi liquid theory. It is expected that long-lived quasiparticles exist near the Fermi surface, which experience decoherence as the energy scale increases \cite{ceMisra2008,ce2Coleman2015}. The lifetime of these quasiparticles is proportional to the inverse of the imaginary part of the self energy on the Matsubara axis, $\tau \propto [Im\{\Sigma_f(i0^+)\}]^{-1}$, where $\tau$ is the quasiparticle lifetime at the Fermi surface and $\Sigma_f(\omega)$ is the local self energy of the $f$-electron states calculated from DMFT. For CeB$_3$C$_3$ between 50 K to 200 K, we predict a convergence towards the Landau-Fermi liquid for the 4$f_{5/2}$ states since $Im\{\Sigma_{5/2}(i0^+)\}$ rapidly approaches zero as temperature is lowered to 50 K as shown in Fig. \ref{fig:imsigtempevo}(a). This indicates a fermi-liquid energy scale of around $T^* \approx 50 K$. The 4$f_{7/2}$ states remain coherent throughout the temperature evolution but contain no weight at the Fermi level. This convergence towards coherence provides evidence for the formation of a Fermi-liquid state at low temperatures. The spectral function at 50 K provides interesting insights into the strong electronic correlations. We observe an almost directional dependence on the Fermi surface, as indicated by the lack of spectral weight at the R point. Additionally, we find that electron hole pockets emerge at the M points in the cubic Brillouin zone. 

\begin{figure*}
    \includegraphics[width=0.9\textwidth]{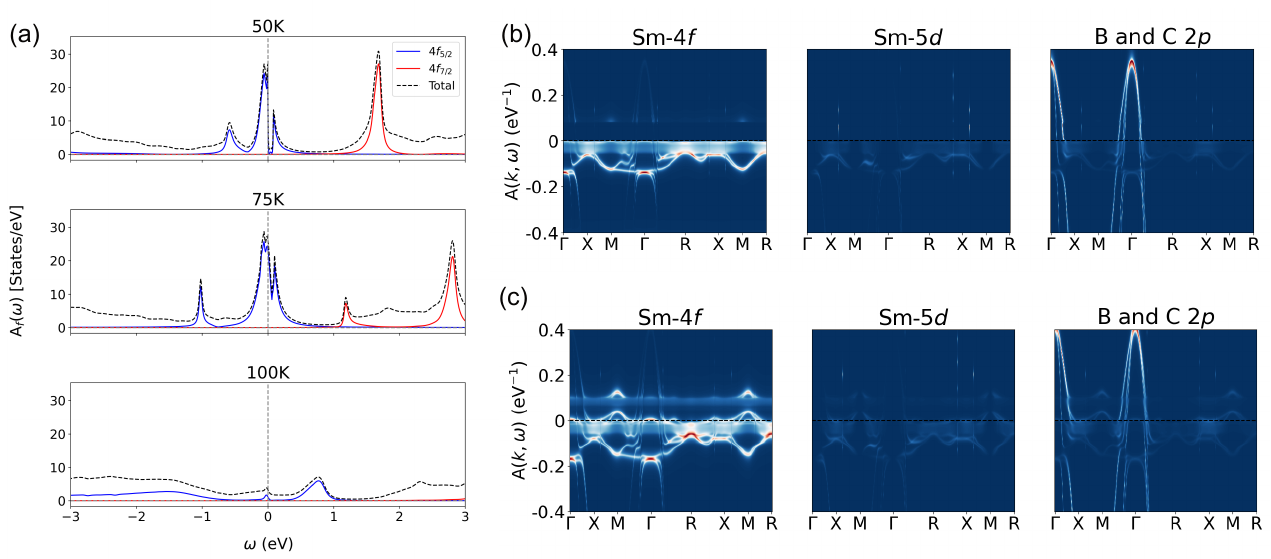}
    \caption{(a) Temperature dependent evolution of the $f$-electron density of states of SmB$_3$C$_3$. Orbital-resolved spectral function for SmB$_3$C$_3$ at (a) 50K and (b) 75K. Dominant hybridization occurs from B and C 2$p$ orbitals. The $y$-axis units are in eV, and intensity is scaled for clarity. }
    \label{fig:Sm}
\end{figure*}
Heavy fermion materials are typically characterized by a substantial linear temperature dependence of specific heat characterized by the Sommerfeld coefficient $\gamma$ at low temperatures \cite{ghfBassani2005}. To estimate this quantity, we use the renormalized density of states. At 50 K, the quasiparticle renormalization for the $f$ states is estimated by \cite{ce2Coleman2015}
\begin{equation}
    Z = \left[ 1- \left.\frac{d Im\{\Sigma_{f,5/2}(i\omega_n)\}}{d(i\omega_n)}\right|_{i\omega_n\rightarrow 0^+} \right]^{-1},
\end{equation}
which yields a value of $Z_{50K, f} = 0.158$. The linear part of the specific heat, $\gamma$, can subsequently be estimated as 
\begin{equation}
    \gamma = \frac{\pi^2 k_B^2}{3}\frac{N^{LDA}(E_f)}{Z_{50K,f}},
\end{equation}
where $N^{LDA}(E_f)=12.342$ $eV^{-1}$ is the density of states at the Fermi level, estimated from the local density approximation (LDA) DFT calculations, and $k_B$ is Boltzmann's constant. We use the density of states from LDA since the temperature of 50 K in DMFT simulations is not low enough to fully capture the transition to a heavy fermion state. Incorporating these values, we predict a linear specific heat coefficient of $\gamma = 183 \frac{mJ}{molK^2}$ which is an order of magnitude above simple metals and in line with concentrated Kondo systems \cite{ce2Bredl1978}.

Sm-based compounds offer another avenue to engineer strong correlations into C-B clathrates. In general, Sm compounds exhibit mysterious properties even within the sphere of strongly correlated materials. Theoretical explanations for small moment magnetism, mixed valency \cite{sm2Ryousuke2018}, and topological Kondo insulating effects \cite{sm2Neupane2013} remain elusive for this category of compounds. Due to the tendency for valence fluctuations and strongly anisotropic magnetic moments \cite{sm2Mahanti2008,sm2McEwen1974}, a clathrate compound containing Sm could unveil new and captivating physics, potentially leading to novel applications. Furthermore, such unique structures may shed light on open questions regarding the role of Sm in strongly correlated materials by providing a new crystal environment. Computed phonon dispersion spectra of SmB$_3$C$_3$ (See supplemental material Fig. S1 (b) \cite{supplementlink}) show no imaginary phonon frequencies, indicating structural stability. Additionally, a negative formation energy at atmospheric pressure (approximately -0.98 eV per formula unit) is promising for future synthesis efforts.

To classify the strong electronic correlations, we first investigate the ground state electronic structure of SmB$_3$C$_3$. At higher temperatures, localized $f$-moment electrons generally disperse within the lower and upper Hubbard bands. Figure \ref{fig:Sm}(a) depicts the transition of the $f$-electron DOS from 100 K to 50 K, where a clear transference of spectral weight towards the Fermi surface is evident, indicating strong Kondo lattice effects. The spectral weight from the wide peaks situated at around -1.5 eV and 0.75 eV steadily transfer towards the Fermi level and sharpens, resulting in localized peaks in the density of states. This suggests that the local moment phase is screened as the temperature decreases. The $f$-level occupancy is n$_f=5.59$ at T = 50 K, in contrast to n$_f=5.12$ at T = 100 K, implying that SmB$_3$C$_3$ enters a mixed valency regime as the temperature drops. The zero frequency imaginary part of the self energy for the 4$f_{5/2}$ states never reaches zero as shown in Fig. \ref{fig:imsigtempevo}(b). In addition, the occupancy of the 4$f_{5/2}$ state is quite large indicating a gapped phase. 

The orbital-resolved spectral functions (Fig. \ref{fig:Sm}(b-c)) indicate that the predominant hybridization contributions in SmB$_3$C$_3$ originates from the B and C $p$ orbitals. As the temperature decreases from 75 K to 50 K, a discernible enhancement in orbital weight near the Fermi level becomes evident for the $p$ orbitals. Concurrently, a strong renormalization of spectral weight occurs below the Fermi level in the Sm $f$ orbitals. The formation of a gap is apparent, similar to the Kondo insulating gap found in black phase of SmS \cite{sm2Kang2015}. The emerging state, primarily dictated by the $j=5/2$ angular momentum states, exhibits a 90 meV gap at 50 K. This supports the hypothesis that SmB$_3$C$_3$ may be a Kondo insulator.

\section{CONCLUSION}

In summary, both clathrates display signs of strong Kondo lattice behavior as temperature is lowered. They are predicted to be dynamically stable and thermodynamically favorable under ambient pressure, making them strong candidates for experimental synthesis. CeB$_3$C$_3$ is predicted to be a potential heavy fermion compound with a Kondo temperature below 50 K. This clathrate displays both Fermi-liquid behavior and concentrated Kondo lattice behavior at intermediate temperatures. 
SmB$_3$C$_3$ appears to transition towards a Kondo insulating state, with a notable redistribution of spectral weight around the Fermi level coupled with a very small band gap. Though the orbital-resolved spectral functions do not unveil topological characteristics at 50 K and ambient pressure, applying pressure may reveal novel aspects of the phase diagram. These observations suggest an intriguing potential for future research.  Moreover, the discovery of the two heavy-fermion clathrates ushers in a new domain of materials that exhibit strong electronic correlation effects, providing a robust platform for investigating such phenomena.

\begin{acknowledgments}
This work was supported by the startup funds of the office of the Dean of SASN of Rutgers University-Newark. The authors acknowledge the Office of Advanced Research Computing (OARC) at Rutgers for providing access to the Amarel cluster and associated research computing resources.
\end{acknowledgments}

\bibliography{biblio}

\end{document}